\documentclass[american,twocolumn,prl]{revtex4-1}
\usepackage[T1]{fontenc}
\usepackage[utf8]{inputenc}
\usepackage[english]{babel}
\usepackage{array}
\usepackage{rotating}
\usepackage{varioref}
\usepackage{booktabs}
\usepackage{units}
\usepackage{multirow}
\usepackage{amsmath}
\usepackage{amssymb}
\usepackage{graphicx}
\usepackage{xcolor}

\usepackage{textgreek}
\usepackage{lineno}
\usepackage{wasysym}
\usepackage{marvosym}
\usepackage{hyperref}

\definecolor{link_color}{HTML}{00406E} 
\definecolor{cite_color}{HTML}{590000} 

\hypersetup{
    colorlinks   = true, 
    urlcolor     = link_color, 
    linkcolor    = black, 
    frenchlinks  = true, 
    citecolor    = cite_color 
}

\makeatletter
\DeclareTextCommand{\textprime}{\encodingdefault}{%
  \mbox{$\m@th'\kern-\scriptspace$}%
}
\makeatother

\usepackage{ upgreek }
\newcommand{\gdp}{\textgamma\textprime\textprime} 
\newcommand{\gp}{\textgamma\textprime} 
\newcommand{\g}{\textgamma} 
\newcommand{\ratio}{$\varepsilon_{3}/\varepsilon_{1}$} 
\newcommand{\fig}{Figure~} %
\newcommand{\tab}{Table~} %
\newcommand{\equ}{Equation~} %
\newcommand{\hexx}{hexagonal\textsubscript{x}} %
\newcommand{\hexy}{hexagonal\textsubscript{y}} %
\newcommand{\mJm}{mJ\,m^{-2}} %
\newcommand{\mv}[1]{\mathbf{#1}}

\makeatother

\begin{document}
\title{Phase-field modeling of  \g{}/\gdp{} microstructure formation in Ni-based
superalloys with high \gdp{} volume fraction}


\author{Felix Schleifer}
\affiliation{Metals and Alloys, University of Bayreuth, Prof.-Rüdiger-Bormann-Str.\,1, 95447 Bayreuth, Bavaria, Germany}
\author{Markus Holzinger}
\affiliation{Metals and Alloys, University of Bayreuth, Prof.-Rüdiger-Bormann-Str.\,1, 95447 Bayreuth, Bavaria, Germany}
\author{Yueh-Yu Lin}
\affiliation{Metals and Alloys, University of Bayreuth, Prof.-Rüdiger-Bormann-Str.\,1, 95447 Bayreuth, Bavaria, Germany}
\author{Uwe Glatzel}
\affiliation{Metals and Alloys, University of Bayreuth, Prof.-Rüdiger-Bormann-Str.\,1, 95447 Bayreuth, Bavaria, Germany}
\author{Michael Fleck}
\email{michael.fleck@uni-bayreuth.de}
\affiliation{Metals and Alloys, University of Bayreuth, Prof.-Rüdiger-Bormann-Str.\,1, 95447 Bayreuth, Bavaria, Germany}

\begin{abstract}
The excellent mechanical properties of the Ni-based superalloy IN718
mainly result from coherent \gdp{} precipitates. Due to a strongly
anisotropic lattice misfit between the matrix and the precipitate
phase, the particles exhibit pronounced plate-shaped morphologies. Using
a phase-field model, we investigate various influencing factors that determine
the equilibrium shapes of \gdp{} precipitates, minimizing the sum
of the total elastic and interfacial energy. Upon increasing precipitate
phase fractions, the model predicts increasingly stronger particle-particle
interactions, leading to shapes with significantly increased aspect
ratios. Matching the a priori unknown interfacial energy density to
fit experimental \gdp{} shapes is sensitive to the phase content
imposed in the underlying model. Considering vanishing phase content
leads to $\unit[30]{\%}$ lower estimates of the interfacial energy
density, as compared to estimates based on realistic phase fractions
of $\unit[12]{\%}$. We consider the periodic arrangement of precipitates
in different hexagonal and rectangular superstructures, which result
from distinct choices of point-symmetric and periodic boundary conditions.
Further, non-volume conserving boundary conditions are implemented
to compensate for strains due to an anisotropic lattice mismatch between
the \g{} matrix and the \gdp{} precipitate. As compared to conventional
boundary conditions, this specifically tailored simulation configuration
does not conflict with the systems periodicity and provides substantially
more realistic total elastic energies at high precipitate volume fractions.
The energetically most favorable superstructure is found to be a hexagonal
precipitate arrangement.

{{}\copyright~ 2020. The manuscript is made available under the license \href{http://creativecommons.org/licenses/by-nc-nd/4.0/}{CC-BY-NC-ND 4.0
\includegraphics[width=1cm]{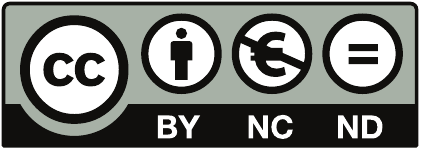}}}

\end{abstract}

\maketitle

\section{Introduction}

\label{sec:Introduction}Ni-based superalloys have various applications
at elevated temperatures, especially in stationary gas turbines and
airplane engines. The main strengthening mechanism, which makes these
alloys applicable for high temperatures is particle strengthening
by coherent precipitations \citep{paulonis1969}. Apart from the most
prominent case, the \gp{} strengthening of Ni-based superalloys for
turbine blade materials, a series of alloys exist that are mainly
strengthened by the tetragonal metastable \gdp{} phase. These Nb-containing
alloys, such as the well-known IN718, can be cast, forged, machined
and welded which renders them ideal candidates for industrial applications.

As IN718 contains up to $\unit[6]{\%}$ of the cubic $\mathrm{L1_{2}}$
phase \gp{}, several authors modified its composition in order to
increase the volume fraction of \gdp{} phase and to get rid of \gp{}
precipitates while maintaining the composition of the matrix \citep{Kirman_1970,Kusabiraki1994,Kusabiraki1996,Kusabiraki1999}.
\tab\ref{tab:composition} shows the nominal composition of IN718
and a derivative IN718M without \gp{} forming elements. Recent
studies aim at deliberate co-precipitation of \gp{} and \gdp{} to
exploit ripening inhibiting effects \citep{Detor2018,Shi2019a} and
a dual lattice microstructure \citep{Mignanelli2017,Mignanelli2018}.
IN718 powders also gain importance for additive manufacturing
in the industrial environment \citep{Amato20122229,Yap2015,StroessnerTerockGlatzel2015,Trosch2016}.
It was recently found that alloys containing only one orientational
variant of \gdp{}, a so-called single-variant microstructure, can
be used to tailor creep resistant materials \citep{Zhang2019}.

\begin{figure}[tbh]
\begin{centering}
\includegraphics[width=\columnwidth]{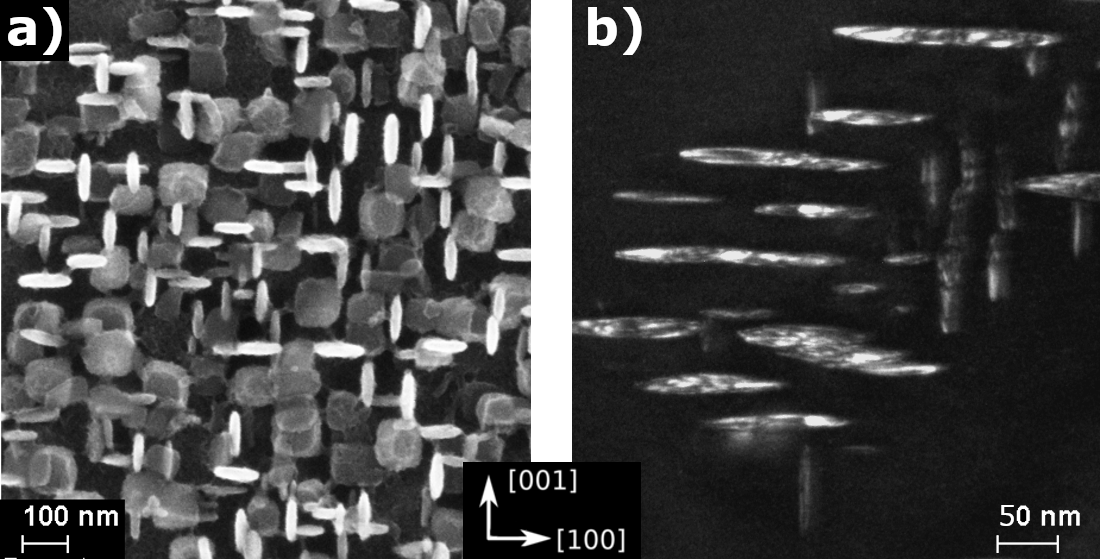}
\par\end{centering}
\caption{a) Scanning electron microscopy (SEM) image of a \g{}/\gdp{} microstructure
in the Nb-rich region of IN718M (composition given in \tab\ref{tab:composition}).
The orientational variants of the \gdp{} precipitates are clearly
distinguishable. b) Dark-field transmission electrop microscopy (TEM)
image of \gdp{} precipitates in IN718M.\label{fig:SEM}}
\end{figure}

\fig\ref{fig:SEM}a) shows a scanning electron microscope (SEM) image
of a \g{}/\gdp{} microstructure in IN718M after homogenization
at $\unit[1423]{K}$ for $\unit[2]{h}$. The precipitates are found
in the Nb-rich interdendritic region. \tab\ref{tab:composition}
also shows the local composition measured by energy-dispersive x-ray
spectroscopy. The precipitates are arranged regularly in three spatial
orientations perpendicular to one another and show a plate-shaped  morphology.
The volumetric \gdp{} phase fraction is $\unit[\geq12]{\%}$. \fig\ref{fig:SEM}b)
shows a dark-field transmission electron microscope (TEM) image of
\gdp{} preciptates in IN718M.

{
\tabcolsep=0.13cm
\begin{table}[tbh]
{\small
\begin{tabular}{rccccccc}
\toprule 
Composition in wt.\ \%  & \multicolumn{1}{c}{Ni} & \multicolumn{1}{c}{Cr} & \multicolumn{1}{c}{Nb} & \multicolumn{1}{c}{Mo} & \multicolumn{1}{c}{Fe} & \multicolumn{1}{c}{Al} & \multicolumn{1}{c}{Ti}\tabularnewline
\midrule
IN718 nominal max. & 55.0 & 21.0 & 5.5 & 3.3 & bal. & 0.8 & 1.2\tabularnewline
IN718 nominal min. & 50.0 & 17.0 & 4.8 & 2.8 & bal. & 0.2 & 0.7\tabularnewline
IN718M nominal & 58.0 & 18.0 & 5.0 & 3.0 & bal. & -- & --\tabularnewline
IN718M measured & 56.8 & 17.6 & 6.6 & 3.2 & 15.8 & -- & --\tabularnewline
\bottomrule
\end{tabular}
}
\caption{Nominal composition in wt.\ \% of IN718 and its derivative
IN718M together with the measured composition in the Nb-rich
region of IN718M as shown in \fig\ref{fig:SEM}a).\label{tab:composition}}
\end{table}
}

The shapes and spatial arrangements of coherent misfitting precipitates
have been modeled assuming a single precipitate embedded in infinite
matrix \citep{khachaturyan1967,khachaturyan1974spatially,Johnson1984,Voorhees1992,Thompson1994,Thompson1999,Li2004}.
The influence of elastic inhomogeneity \citep{schmidt1997equilibrium}
and periodic arrangements of precipitates with higher precipitate
volume fractions \citep{Mueller2000} on precipitate shapes were studied
using the boundary integral method. Comparison of simulated precipitate
shapes to experimentally observed microstructures can be used to obtain
realistic values of the interfacial energy density \citep{lanteri1986morphology,Devaux2008117,Holzinger2019}.

The phase-field method is widely considered to be a powerful tool for modeling
solidification as well as solid-state phase transformations based
on a diffuse description of the phase boundaries \citep{Chen2002,AstaBeckermKarma012009,Steinba2009,WangLi012010,DeWittThornton2018}.
Consistent phase-field modeling of precipitation bases on a thermodynamic
functional, which is an integral over a local phenomenological potential
energy density. This phenomenological energy density may take into
account phase boundary energy, multi-component solute distribution
and elastic strain caused by external loads or lattice misfit between
the phases. By general variational principles, one can subsequently
derive a consistent set of coupled partial differential equations
that describes the kinetics of the microstructure evolution in the
chosen configuration. The phase-field method is frequently used to study
the temporal evolution of \g{}/\gp{} microstructures \citep{wang1998field,zhu2004three,GaubertLe-BouaFinel012010,MushongeraFleckKundinWangEmmerich2015,PangLiWuLiuHou2015,BhaskarPhasefield2018}
as well as \g{}/\gdp{} systems \citep{Zhou2014270,JiLouRowattZhangSimpsonChen2016}.
The variational formulation of the phase-field method makes it a useful
tool to determine equilibrium precipitate morphologies \citep{wang1991shape,wang1993kinetics,P.H.Leo,Cottura2015,Jokisaari2017,Bhadak2018}
and precipitate interactions \citep{degeiter2020}. In the scope of
this work, we develop a phase-field model to extensively evaluate factors
influencing two-dimensional \gdp{} precipitate shapes. We consider anisotropic
and phase-dependent elastic properties, a tetragonally anisotropic
misfit, elastic particle interactions at high volume content as well
as an isotropic energy density of the \g{}/\gdp{} interface.

\section{The \g{}/\gdp{} microstructure in Ni-based superalloys\label{sec:The-gammaDP-phase}}

The ordered $\mathrm{D}0_{22}$ phase \gdp{} has a body-centered
tetragonal (bct) crystallographic structure and is of the stoichiometric
type Ni$_{3}$Nb. In an fcc matrix, it forms distinctive plate-shaped
 precipitates. The face-centered cubic (fcc) unit cell of the matrix
phase \g{} can be described by a single lattice parameter $a_{\gamma}$
that is the distance between two atom sites along a $\langle100\rangle$
direction. The bct unit cell of the \gdp{} phase can be made up by
two fcc unit cells stacked on top of each other with the central atom
site and the corner atom sites being Nb atoms. One can now distinguish
two lattice parameters $a_{\gamma''}$ and $c_{\gamma''}$, the latter
being also referred to as the tetragonal axis. The plate normal of
the precipitate is always parallel to the $c_{\gamma''}$ direction.

Coherent precipitation of \gdp{} in a \g{} matrix is possible due
to the relations 
\begin{equation}
\begin{array}{cccc}
a_{\gamma''} & \approx & a_{\gamma}\\
c_{\gamma''} & \approx & 2a_{\gamma} & ,
\end{array}\label{eq:coherency}
\end{equation}
when
\begin{equation}
\ensuremath{\langle}100\rangle_{\gamma}\parallel\ensuremath{\langle}100\rangle_{\gamma''}.\label{eq:coherency-1}
\end{equation}
This means that there are three possible orientational variants in
which the tetragonal phase can coherently precipitate (see \fig\ref{fig:SEM})
\citep{SlamaAbdellaoui_2000}. Note that \equ(\ref{eq:coherency})
states that there is a misfit of the lattice parameters that leads
to strains when lattice coherency is kept. Two distinguishable misfit
strains $\varepsilon_{1}$ and $\varepsilon_{3}$ in $a_{\gamma''}$
and $c_{\gamma''}$ direction, respectively, can be found.

\begin{figure}[tbh]
\begin{centering}
\includegraphics[width=\columnwidth]{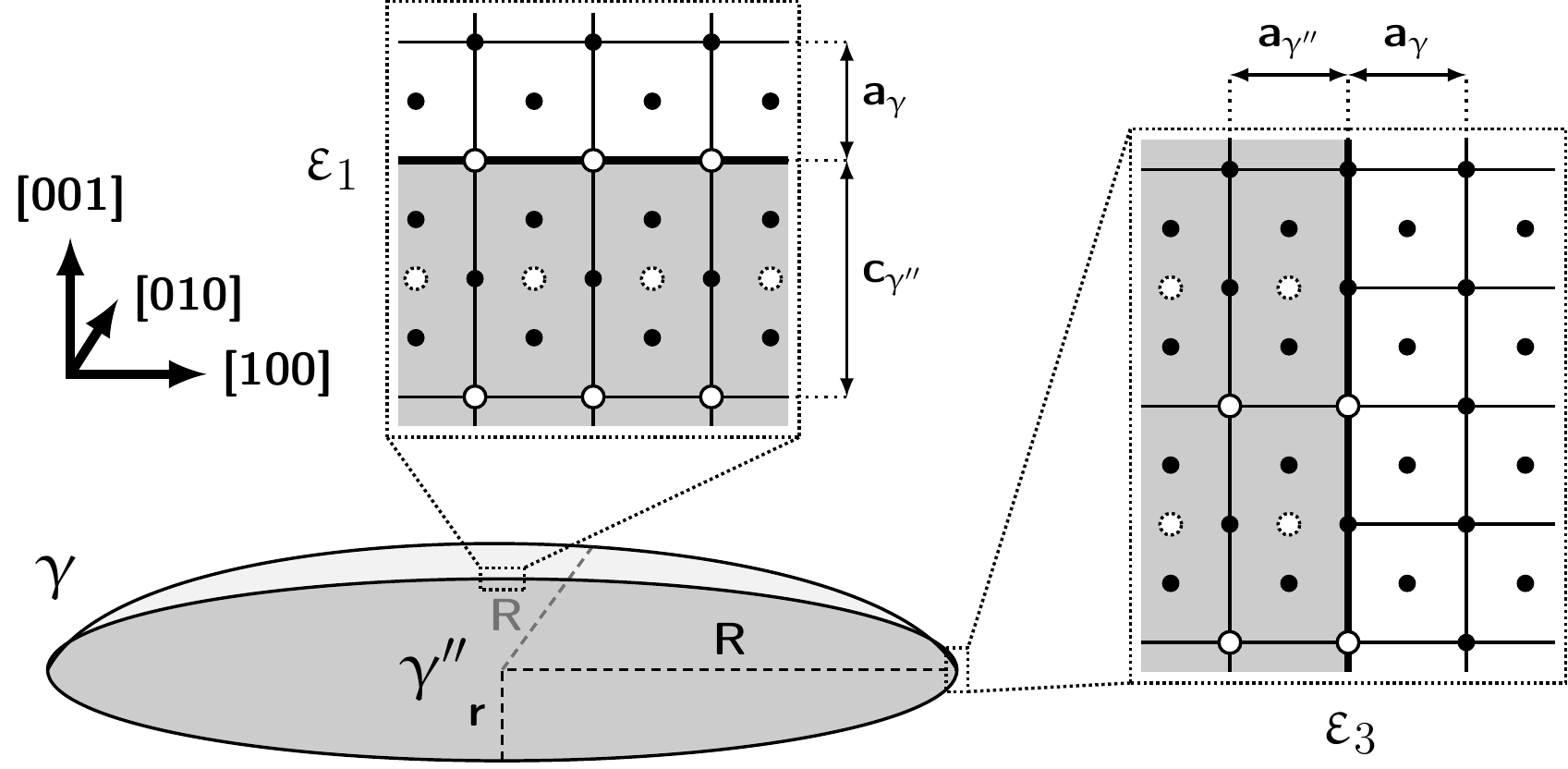}
\par\end{centering}
\caption{Cut through the center of a \gdp{} precipitate (gray) parallel to
the $\left(010\right)$ plane. The two major half axes $R$ and the
minor half axis $r$ of the precipitate are indicated by dashed lines.
The two orientations of the respective unit cells in the interface
are shown that lead to two distinguishable misfit strains $\varepsilon_{1}$
and $\varepsilon_{3}$. Black circles indicate Ni atoms and white
ones Nb, respectively. Dotted Nb atoms lie below the cut plane. The
lattice parameters $a_{\gamma}$ ,$a_{\gamma^{''}}$ and $c_{\gamma^{''}}$
are also indicated. \label{fig:precipitates_scheme}}
\end{figure}

\fig\ref{fig:precipitates_scheme} shows a schematic drawing of a
\gdp{} precipitate. The precipitate is depicted as an oblate spheroid
with two major and a minor half axis $R$ and $r$, respectively.
The orientational relations given in \equ(\ref{eq:coherency-1})
are depicted at the interface. To experimentally quantify the shape
of \gdp{} precipitates one usually takes the aspect ratio that is
defined as the ratio $R/r$ or, for non-elliptical precipitates the
ratio of the plate diameter to its thickness. Usually, \gdp{} precipitates
exhibit major radii of less than $\unit[130]{nm}$ and reportedly
start to lose full coherency at major radii larger than $\unit[25]{nm}$
\citep{Cozar1973a,slama1997,Devaux2008117,JiLouRowattZhangSimpsonChen2016}.

\subsection{Elastic constants\label{subsec:Elastic-material-parameters}}

General elastic behavior is described by the tensor of elasticity
$C_{ijkl}$. In this work, we use the tensor of elasticity in Voigt
notation $C_{ij}$. This is a reduction to a $6\times6$ tensor that
is not invariant under rotation. The cubic \g{} and the tetragonal
\gdp{} phase both exhibit anisotropic elastic properties due to their
crystallographic symmetry. A set of three independent elastic constants
is needed to describe cubic anisotropy ($C_{11}$, $C_{12}$, $C_{44}$)
and six are needed for tetragonal anisotropy, respectively ($C_{11}$,
$C_{33}$, $C_{12}$, $C_{13}$, $C_{44}$, $C_{66}$).  Isotropic
elasticity is described by a set of two independent constants with
$C_{11}=2C_{44}+C_{12}$.

In this work we use experimental data generated by resonance ultrasound
spectroscopy (RUS) for polycrystal (isotropic) and single crystal
(anisotropic) samples of IN718. We assume that matrix and precipitates
in IN718 and IN718M have comparable elastic properties.
For the \gdp{} phase data from a first-principles study was used
\citep{Connetable2011,Dai2010414}. This data is valid at $\unit[0]{K}$,
therefore, an estimate about temperature dependence has to be made.
Due to the lack of experimental data, we propose to assume the same
temperature dependence for all elastic constants considered, as it
was already done by Moore et al.~\citep{Moore2016}. In this work,
we assume a linear temperature dependence of the elastic constants
\begin{align}
C_{ij}(T)=C_{ij}^{0K}(1-\beta\varDelta T) & ,\label{eq:T_dependence}
\end{align}
where $T$ is the temperature, $C_{ij}^{0K}$ is an elastic constant
at $\unit[0]{K}$ and $\beta$ is the coefficient of temperature dependence.
To determine $\beta$, experimental data for pure Nickel from $\unit[0]{K}$
to room temperature is used \citep{Simmons1965,Luo2011}. For higher
temperatures, we assume the same temperature dependence as it was
determined for a \gp{} single crystal (Ni-Al23-Ti1-Ta1) via RUS \citep{Fleck2018}.
We find $\beta$ to be $\unit[2.3\cdot10^{-4}]{K^{-1}}$. \tab\ref{tab:Elastic-constants}
shows the extrapolated elastic constants of \gdp{} at $\unit[998]{K}$.
Elastic homogeneity is given when matrix and precipitate have the
same tensor of elasticity ($C_{ij}^{\gamma}=C_{ij}^{\gamma''}$).

The elastic constants of a single crystal alloy composed of $N$ coherent
phases can be estimated by a simple rule of mixture
\begin{align}
\bar{C}_{ij} & =\sum^{\alpha}C_{ij}^{\alpha}V_{\alpha},\label{eq:mixture_rule}
\end{align}
where $\bar{C}_{ij}$ is an elastic constant of the alloy, $C_{ij}^{\alpha}$
is the elastic constant of phase $\alpha$ and $V_{\alpha}$ is the
volume fraction of $\alpha$. Knowing the elastic properties for all
but one phase and for the alloy itself, it is possible to estimate
the tensor of elasticity for the remaining phase. In the case of \gdp{}
precipitates in a cubic matrix, the three orientational variants appear
statistically. The tensor of elasticity of such an alloy maintains
cubic symmetry. To find the tensor of elasticity, with cubic symmetry,
for a mixture of all three orientation variants $\bar{C}_{ij}^{\gamma''}$
of the \gdp{} phase, we apply the modified mixture rule
\begin{align}
\bar{C}_{11}^{\gamma''}&=\left(2C_{11}^{\gamma''}+C_{33}^{\gamma''}\right)/3\\
\bar{C}_{12}^{\gamma''}&=\left(C_{12}^{\gamma''}+2C_{13}^{\gamma''}\right)/3\\
\bar{C}_{44}^{\gamma''}&=\left(2C_{44}^{\gamma''}+C_{66}^{\gamma''}\right)/3.
\label{eq:gamma_dp_mixture}
\end{align}
Having estimated the complete tensors of elasticity for the IN718M
alloy and the \gdp{} phase, the anisotropic tensor of elasticity
of the matrix can be calculated using \equ(\ref{eq:mixture_rule})
and (\ref{eq:gamma_dp_mixture}). In \tab\ref{tab:Elastic-constants}
also a stiffness contrast $C_{ij}^{\gamma''}/C_{ij}^{\gamma}$ is
given. In the case of anisotropic elasticity, we apply the mixture
rule in \equ(\ref{eq:gamma_dp_mixture}) to compare the respective
tensors of elasticity. We report a mean value as the different elastic
constants show different stiffness contrasts.

To evaluate the influence of inhomogeneous stiffness of the phases
and of crystallographic anisotropy on the precipitate shapes, we use
different sets of elastic constants. Phase-independent isotropic elastic
properties leave only the anisotropic lattice misfit to determine
the precipitate shape. Phase-wise isotropic data allows to observe
the influence of elastic inhomogeneity. Individual cubic and tetragonally
anisotropic phase data describes the elastic properties of the system
in the most comprehensive way.

{
\tabcolsep=0.13cm
\begin{table}[tbh]
{\small
\begin{tabular}{crccccccr}
\toprule 
 &  &  \multicolumn{6}{c}{elastic constants in GPa} & {contrast } \tabularnewline
 &  &$C_{11}$&$C_{33}$&$C_{12}$&$C_{13}$&$C_{44}$&$C_{66}$ & {$C_{ij}^{\gamma''}/C_{ij}^{\gamma}$}\tabularnewline
\midrule
\multirow{3}{*}{\vspace{-4mm}\begin{turn}{90}
isotropic
\end{turn}}  & homogen.$^{1}$ &(242)&  &120&  &61&  & $1.0$ \rule[-3mm]{0mm}{3mm} \tabularnewline
 &matrix$^{3}$ &(240)&  &120&  &60& \rule[-3mm]{0mm}{3mm} & \multirow{1}{1.2cm}{$\Bigg\}<1.1$}\tabularnewline
 &\gdp{} phase$^{4}$ &(250) &  & 120 &  & 65 &  \rule[-6mm]{0mm}{3mm} &  \tabularnewline
\multirow{3}{*}{ \vspace{-5mm} \begin{turn}{90}
anisotropic
\end{turn}}  & homogen.$^{1}$ & 205 &  & 145 &  & 90 &  &$1.0$ \rule[-3mm]{0mm}{3mm} \tabularnewline
 &matrix$^{3}$ &200&  &150&  &90 &  \rule[-3mm]{0mm}{3mm} & \multirow{1}{1.2cm}{$\Bigg\}\approx0.9$}\tabularnewline
 &\gdp{} phase$^{2}$ &220&240&140&120& 88 & 87 \rule[-2mm]{0mm}{3mm} & \tabularnewline
\bottomrule
\end{tabular}
}

\caption{Elastic constants at $\unit[998]{K}$ used in this work. Data for
the individual phases and for an elastically homogeneous system ($C_{ij}^{\gamma}=C_{ij}^{\gamma''}$)
are provided as well as isotropic and anisotropic data sets to evaluate
respective effects separately. Values of a mean stiffness contrast
$C_{ij}^{\gamma''}/C_{ij}^{\gamma}$ are provided. Data is generated
via: $^{1}$resonance ultrasound spectroscopy $^{2}$first-principles
calculations \citep{Connetable2011,Dai2010414} and \equ(\ref{eq:T_dependence})
$^{3}$\equ(\ref{eq:mixture_rule}) $^{4}$\equ(\ref{eq:gamma_dp_mixture})
\label{tab:Elastic-constants}}
\end{table}
}
All evaluations of the elastic moduli of the phases based \equ(\ref{eq:mixture_rule})
are dependent on the volume fraction $V_{\alpha}$. In this work,
we assume a \gdp{} volume fraction of $\unit[12]{\%}$ which is in
accordance with previously reported values \citep{brooks.88,Theska2018}
and higher than it was calculated using the Thermo-Calc software with
the TCNi8 database.

\subsection{Anisotropic lattice misfit \label{subsec:misfit}}

Due to the two distinguishable lattice parameters $a$ and $c$ in
\gdp{} and the strict orientation relation of the coherent interface
in a \g{}/\gdp{} microstructure the stress-free transformation strain
$\varepsilon_{ij}^{0}$ exhibits a tetragonal symmetry. In Voigt notation,
this misfit strain tensor $\varepsilon_{0}$ of the \gdp{} phase
is
\begin{align}
\varepsilon_{0} & =\left(\begin{array}{ccc}
\varepsilon_{1}, & \varepsilon_{1}, & \varepsilon_{3}\end{array},\,0,\,0,\,0\right)^{T},\label{eq:misfit_strain}
\end{align}
with the two distinguishable dilatational misfit strains $\varepsilon_{1}$
and $\varepsilon_{3}$. Those are calculated from the experimentally
determined lattice parameters of the phases via
\begin{align}
& \varepsilon_{1}= \frac{a_{\gamma''}-a_{\gamma}}{a_{\gamma}}, \qquad
\varepsilon_{3}= \frac{c_{\gamma''}-2a_{\gamma}}{2a_{\gamma}}.
\label{eq:misfit_strain_2}
\end{align}
Note that in the denominator we find only the lattice parameter of
the matrix. This is because we define the eigenstrain of the matrix
to be zero. The lattice parameters of an IN718 matrix and of
the \gdp{} precipitates were measured by Slama et al. \citep{slama1997,SlamaAbdellaoui_2000}
using X-ray diffractometry (XRD) for samples aged at 953\,K and 1023\,K
after quenching. Lattice parameters were also measured using neutron
diffraction \citep{Lawitzki2019}. Measured lattice parameters and
corresponding misfit strains are given in \tab\ref{tab:misfits}.
Other measurements of the lattice parameters in IN718 show similar
misfits \citep{cozar.73,Cozar1973a,OblakPauloniDuvall011974,chaturvedi.83,Kusabiraki1996,Zhang2018}.
Note that the absolute value of $\varepsilon_{3}$ is high compared
to \gp{} precipitates ($\varepsilon_{0}^{\gamma'}\thickapprox-1\ldots-3\cdot10^{-3}$
\citep{VoelklGlatzelFellerKniepmeier1998}) but similar for both XRD
and neutron diffraction. On the other hand, the values of $\varepsilon_{1}$
differ strongly. The misfit ratio \ratio{} is one magnitude larger
for neutron diffraction measurements than for the XRD measurements.

Both measurements, however, determined the in-situ constrained misfit
that is the lattice misfit superimposed by elastic deformation of
the material. A more suiting input parameter would be the unconstrained
misfit that is the misfit calculated from lattice parameters of strain-free
bulk samples of the phases. The unconstrained misfit can notably differ
from the constrained misfit \citep{Voelkl.98,MuellerGlatzelFeller-Kniepmeier1993}.

{
\tabcolsep=0.2cm
\begin{table}[tbh]
{\small
\begin{tabular}{ccccccccc}
\toprule 
 & \multicolumn{3}{c}{lattice param.~in pm} & & \multicolumn{2}{c}{misfit in $10^{-3}$} & \tabularnewline
source & $a_{\gamma}$ & $a_{\gamma''}$ & $c_{\gamma''}$ & & $\varepsilon_{1}$ & $\varepsilon_{3}$ & {\ratio{}}\tabularnewline
\midrule
\citep{slama1997,SlamaAbdellaoui_2000} & 359.5 & 361.4 & 742.1 & & 5.29 & 32.1 & 6.1 \tabularnewline
\citep{Lawitzki2019} & 359.8 & 360.0 & 743.8 & & 0.56 & 33.6 & 60.0 \tabularnewline
\bottomrule
\end{tabular}
}
\caption{Lattice parameters from literature for an IN718 matrix and \gdp{}
precipitates. Misfit strains are given according to \equ(\ref{eq:misfit_strain_2})
as well as the respective misfit ratio \ratio{}.\label{tab:misfits}}
\end{table}
}
\section{Phase-field Modeling of equilibrium precipitate shapes}

\label{sec:Phasefieldmodel} The phase-field model for the simulation of
solid-phase precipitation in multicomponent alloys is based on the
following phenomenological potential functional \citep{Plapp092011,MushongeraFleckKundinWangEmmerich2015,MushongFleckQuerfur032015},
\begin{equation}
\Omega=\int_{V}\omega\left(\varphi,\left\{ \partial_{i}\varphi\right\} ,\left\{ \partial_{k}u_{i}\right\} \right)dV,\label{eq:Grand-Potential-Functional}
\end{equation}
where $\partial_{i}$ denotes an abbreviation for the partial derivative
with respect to the spacial directions $i=x,y,z$, e.g.~$\partial_{x}\equiv\partial/\partial x$.
The potential density $\omega$ splits into an interfacial, a bulk
elastic contribution and a chemical contribution $\omega=\omega_{\mathrm{int}}+\omega_{\mathrm{el}}+\omega_{\mathrm{ch}}$.
The continuous fields describing the evolution of the system are the
phase-field $\varphi$, which discriminates between the fcc matrix ($\varphi=0$)
and the ordered \gdp{} phase ($\varphi=1$) as well as the elastic
displacement field $\left\{ u_{i}\right\} =\left(u_{x},u_{y},u_{z}\right)$,
which describes the local distortion of a material point by elastic
deformations. 

\subsection{Interfacial contribution}

The interfacial contribution to the potential functional is 
\begin{align}
\omega_{\mathrm{int}} & =\frac{\Gamma\xi}{\Gamma_{0}}\left(\partial_{i}\varphi\partial_{i}\varphi\right)+\frac{\Gamma}{\xi\Gamma_{0}}p(\varphi),\label{eq:two-phase-interfacial-energy-density}
\end{align}
where a summation over repeated indices is implied. The phase-field parameter
$\xi$ determines the interface width and the parameter $\Gamma$
corresponds to the interfacial energy density. $\Gamma_{0}$ denotes
a calibration factor for the interface energy density, which is calculated
via the line integral, $\Gamma_{0}=\int\omega_{\mathrm{int}}(\varphi_{0},\left\{ \partial_{i}\varphi_{0}\right\} )\mathrm{d}n/\Gamma$,
where $n$ denotes the direction normal to the interface, and $\varphi_{0}$
is a phase-field with just one full transition from $\varphi=0$ to $\varphi=1$.
Further, the equilibrium potential is 
\begin{align}
p(\varphi) & =\frac{\xi^{2}}{\Delta x^{2}}\left\{ \varphi\left(1-\varphi\right)+\frac{1-a^{2}}{4a^{2}}\log\left(\frac{1-a^{2}}{1-a^{2}\left(1-2\varphi\right)^{2}}\right)\right\} ,\label{eq:equilibrium-potential}
\end{align}
where the parameter $a=\tanh\left(2\Delta x/\xi\right)$ couples
to the discretization grid via the numerical grid spacing $\Delta x$.
This potential has two local minima at $\varphi=0$ and $\varphi=1$,
which correspond to the two distinct phases of the system. In the
continuum limit $\Delta x\rightarrow0$, this potential converges
to the usual quartic double-well potential $p_{\mathrm{\infty}}(\varphi)=8\varphi^{2}(1-\varphi)^{2}$,
as $\lim_{\Delta x\rightarrow0}\xi^{2}a^{2}/\Delta x^{2}=4$.

\begin{figure}[tbh]
\begin{centering}
\includegraphics[width=\columnwidth]{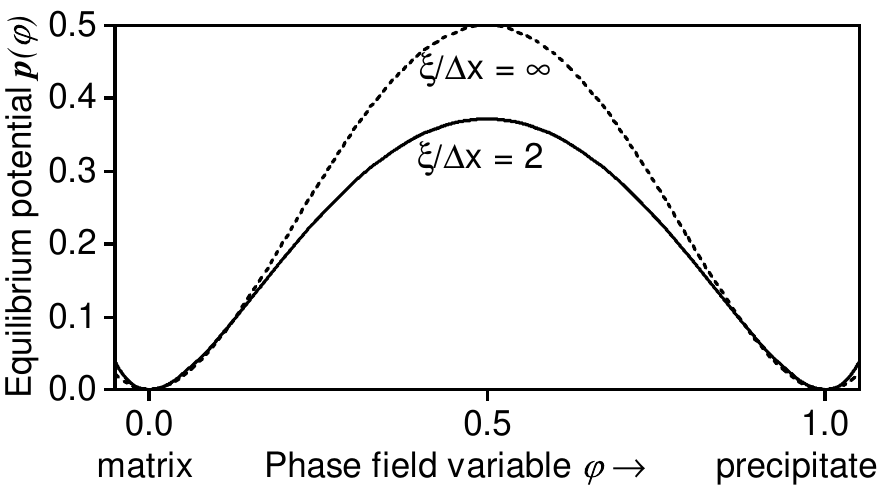}
\par\end{centering}
\caption{\label{fig:equilibrium-potential}Plot of the equilibrium potential
given in \equ(\ref{fig:equilibrium-potential}) as a function of
the phase-field variable $\varphi$ for different values of $\xi/\Delta x$.}
\end{figure}

\fig\ref{fig:equilibrium-potential} shows the equilibrium potential
for $\xi/\Delta x=2$, which will be used throughout this work as
well as for the continuum limit $\xi/\Delta x\rightarrow\infty$.
The reason for choosing this potential is to diminish effects from
the numerical grid in the phase-field model , according to the ideas around
the so-called ``Sharp phase-field method'' as proposed by Finel et al.
\citep{FinelLeBouarDabasAppolairYamada2018,Fleck2019}.

In conjunction with the formulation of the interfacial contribution,
we impose an interpolation function $h(\varphi)=\varphi^{2}(3-2\varphi)$
for any elastic and chemical energy density contribution: $\omega_{\mathrm{el}}+\omega_{\mathrm{ch}}=\omega_{\mathrm{el}}\left(h(\varphi)\right)+\omega_{\mathrm{ch}}\left(h(\varphi)\right)$.
This is the minimal polynomial satisfying the necessary interpolation
conditions $h(0)=0$ and $h(1)=1$ and also having a vanishing slope
at $\varphi=0$ and $\varphi=1$, to not shift the bulk states in
the presence of finite driving forces \citep{Kassner2001,FleckBrenerSpatsch2010}.
The Allen Cahn-type phase-field evolution equation is 
\begin{align}
\frac{\partial\varphi}{\partial t} & =-\frac{2M}{3\Gamma\xi}\frac{\delta\Omega}{\delta\varphi}=\frac{2M}{3\Gamma\xi}\left(\partial_{i}\frac{\partial\omega}{\partial\left(\partial_{i}\varphi\right)}-\frac{\partial\omega}{\partial\varphi}\right)\nonumber \\
 & =M\left(\frac{2}{3\Gamma_{0}}\left\{ \left(\partial_{i}\partial_{i}\right)\varphi-\frac{1}{\xi^{2}}\frac{\partial p(\varphi)}{\partial\varphi}\right\} -\frac{2}{3\Gamma\xi}\frac{\partial h}{\partial\varphi}\left(\frac{\partial\omega_{\mathrm{el}}}{\partial h}\right)\right),\label{eq:Two-phase-field-equation-of-motion}
\end{align}
where $M$ is the interface mobility that is chosen as high as possible
while still guaranteeing stability of the solver. All these equations
are solved by finite difference schemes operating on one fixed square
grid with an explicit Euler-type time integration.

\subsection{Elastic contribution}

The elastic contribution to the potential energy density $\omega_{el}$
is defined in terms of the phase-dependent elastic properties: the
total strain field $\varepsilon$ that is derived from the local displacements
$u_{i}$, the misfit strain $\varepsilon_{0}$ of the precipitate
phase, and the phase-dependent elastic constants $\bar{C}$ \citep{Fleck2018}.
Specifically the elastic energy density is written as
\begin{equation}
\omega_{\mathrm{el}}\left(\varphi,x\right)=\frac{1}{2}\left(\varepsilon-\bar{\varepsilon}_{0}\right)\bar{C}\left(\varepsilon-\bar{\varepsilon}_{0}\right).\label{eq:elasticfreeenergyinindividaulphase-1}
\end{equation}
The phase dependent elastic constants $\bar{C}_{ij}$ as well as the
eigenstrains $\bar{\varepsilon}_{0,i}$ are interpolated as
\begin{align}
\bar{C}_{ij}(\varphi) & =\sum^{\alpha}h(\varphi)C_{ij}^{\alpha}\label{eq:elastic constants-1}\\
\bar{\varepsilon}_{0,i}(\varphi) & =\sum^{\alpha}h(\varphi)\varepsilon\,_{0,i}^{\alpha},\label{eq:Eigenstrain-1}
\end{align}
where $\alpha$ denotes the considered phase and $\varphi$ denotes
the local phase-field parameter. We interpolate on the level of the elastic
parameters, not on the level of the bulk energies to minimize the
amount of interfacial excess energy \citep{DurgaWollantsMoelans2013}.
A Lagrangian formulation in the small strain approximation is used.
The local stress tensor is defined as the partial derivative of the
elastic energy density \equ{}(\ref{eq:elasticfreeenergyinindividaulphase-1})
with respect to the strain tensor,
\begin{align}
\sigma & =\frac{\partial\omega_{el}}{\partial\varepsilon}=\bar{C}\left(\varepsilon-\bar{\varepsilon}_{0}\right).\label{eq:Stress-tensor-1}
\end{align}
The solution for the elastic displacement field is given by the condition
of mechanic equilibrium
\begin{align}
\frac{\delta\Omega}{\delta u_{i}} & =\frac{\partial}{\partial x_{i}}\frac{\partial\omega_{el}}{\partial\varepsilon}=\frac{\partial\sigma}{\partial x_{i}}=0,\label{eq:DLG-mechanical-equilibrium-1-1}
\end{align}
In summary, the model requires to solve the full set of coupled partial
differential equations of second order, as given by \equ{}(\ref{eq:Two-phase-field-equation-of-motion})
for the phase-field and \equ{}(\ref{eq:DLG-mechanical-equilibrium-1-1})
for the elastic displacements. With regard to the mechanical equilibrium,
that is assumed at every time step of the phase-field solver, we perform
a Jacobi relaxation.

\subsection{Chemical contribution}

Realistic microstructure exhibit conserved phase volumes, due to
the conservation of mass \citep{Fleck2018}. Here, the preserved phase
volume is achieved by the chemical contribution to the potential energy
density 
\begin{align}
\omega_{\mathit{\mathrm{ch}}}(t) & =h(\varphi)f_{\mathit{\mathrm{ch}}}(t),\label{eq:PFM-volume-preservation-contribution}
\end{align}
which contains an extra time-de\-pen\-dent and homogeneous driving
force contribution $f_{\mathit{\mathrm{ch}}}(t)$, such that a volume
change of the precipitate phase is prohibited \citep{FleckMushongPilipen102011}.
A more sophisticated description of the kinetics of diffusion-limited
precipitation, which explicitly involves the chemical diffusion of
multiple alloying elements, is also possible \citep{MushongFleckQuerfur032015,Fleck2018},
but beyond the scope of the present work.

The phase fraction $V_{\alpha}$ is conserved when
\begin{align}
0 & =\frac{d}{dt}V_{\alpha}(t)=\int_{V}\frac{\partial}{\partial t}\varphi(\mathbf{x},t)dV.\label{eq:PFM-volume-preservation}
\end{align}
Inserting the right-hand side of phase-field \equ(\ref{eq:Two-phase-field-equation-of-motion})
into (\ref{eq:PFM-volume-preservation}), we obtain the time-dependent
homogeneous driving force 
\begin{align}
f_{\mathit{\mathrm{ch}}}(t)= & \frac{3\Gamma\xi}{2}\frac{G(t)}{H(t)},\label{eq:PFM-volume-pres-chem-pot}
\end{align}
where the following abbreviations are introduced: 
\begin{align}
G(t) & =\int_{V}\left(\partial_{i}(\partial_{i}\varphi)-\frac{8}{\xi^{2}}\frac{\partial p(\varphi)}{\partial\varphi}-\frac{2}{3\Gamma\xi}\frac{\partial\omega_{\mathit{\mathrm{el}}}}{\partial\varphi}\right)dV,\label{eq:PFM-Coeff-for-volume-pres}\\
H(t) & =\int_{V}\frac{\partial h}{\partial\varphi}dV.
\end{align}
 This method for achieving preserved phase volumes can also be applied
to configurations involving more than two phases \citep{NestlerWendlerSelzer072008}.

\subsection{Boundary conditions}

All simulations are carried out in rectangular two-dimensional domains. Elastic
interactions between neighboring precipitates can be adjusted by
changing the domain size with respect to the size of the initial precipitate.
For both cubic and tetragonal symmetry, it is sufficient to model
one quarter of the particle with respective mirror boundary conditions.
Therefore, for every considered spatial dimension exists one mirror
boundary and one opposing domain boundary that ensures periodicity
(henceforth referred to as the periodic boundary). The phase-field is imposed
with no-flux boundary conditions in both cases.

For the elastic displacement fields, standard boundary conditions
such as periodic conditions, strain-free conditions or stress-free
boundary conditions are not useful. Here, a distinction is required
due to the following reasons. If simple stress-free condition are
imposed the state of deformation turns non-uniform, and the boundary
looses its flatness, which conflicts with the periodicity of the system.
Periodic or strain-free conditions conserve periodicity, but do not
 allow for the misfitting precipitate to change the volume of the
simulation domain and/or its aspect ratio, leading to unrealistically
high total deformation energies. 

Therefore, at the mirror boundaries vanishing normal displacements
$u_{n}=0$ and vanishing shear strains $\partial u_{t}/\partial n=0$
are imposed, where $\mv{u}$ denotes the elastic displacement field,
$n$ and $t$ denote the direction normal and tangential to the boundary,
respectively. At the opposing periodic boundary again vanishing shear
strain conditions are imposed. But for the normal displacement, we
homogeneously impose $u_{n}=\bar{u}_{n}$, where $\bar{u}_{n}$ denotes
the average value of all normal displacements at the boundary. Homogeneous
normal displacements are required to avoid conflicts with the periodicity
of the systems and to not introduce strain artifacts. The average
overall normal displacements at the boundary is used to allow for
the volume change, which leads to more realistic total deformation
energies. This boundary condition has previously been applied to finite-element
modeling of misfit stresses \citep{Glat1989,Probst-Hein1999,Preussner2005,Pollock2012}
and equilibrium shapes \citep{Mueller2000} in the \g{}/\gp{} system.
 Due to these boundary conditions, all opposing domain boundaries
stay parallel yet no spurious elastic energy is introduced through
artificial volume conservation.

\begin{figure*}[tbh]
\begin{centering}
\includegraphics[width=2\columnwidth]{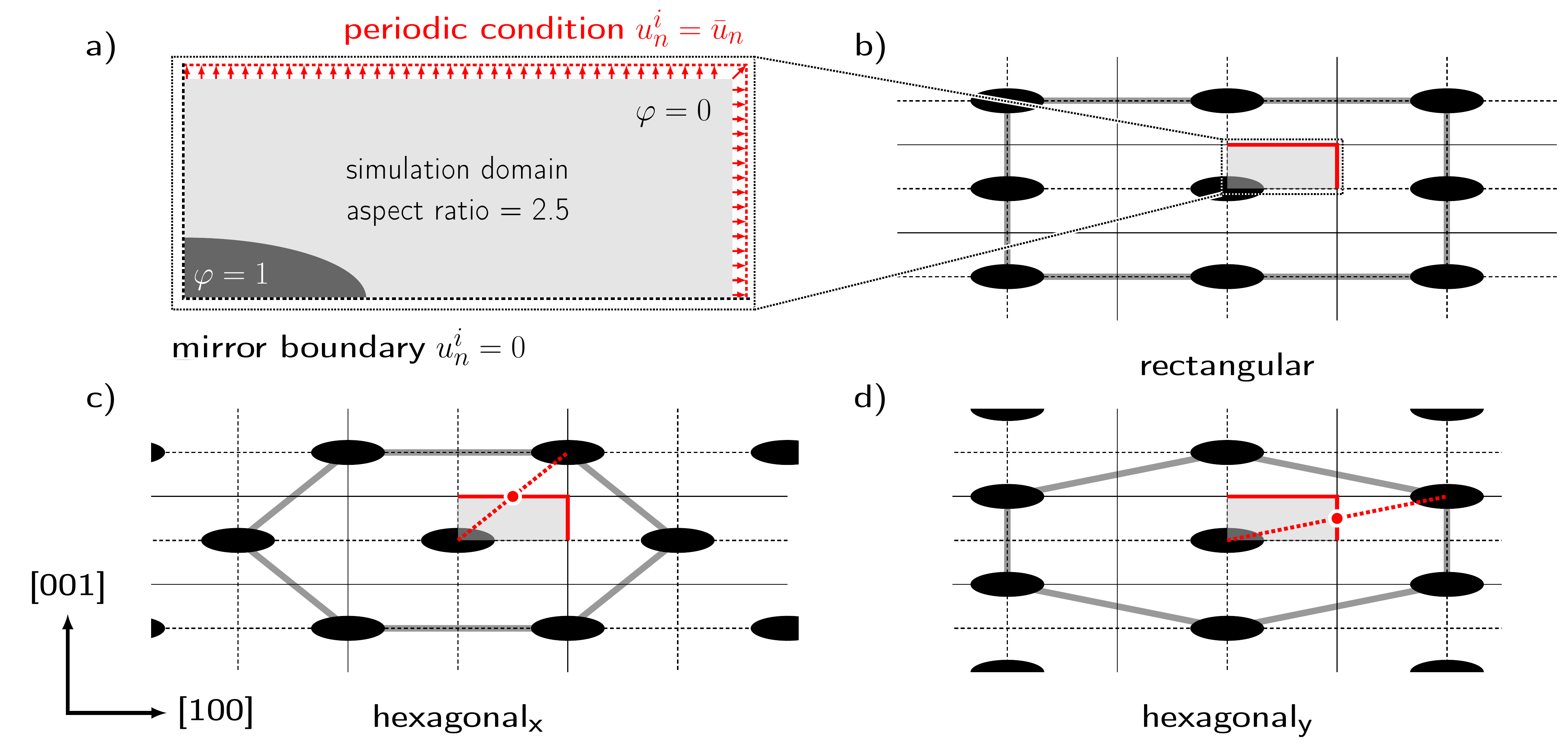}
\par\end{centering}
\caption{a) Rectangular simulation domain. The specifically tailored boundary
conditions for the displacements normal to the boundaries are indicated
in red. Mirror boundaries are indicated by dashed lines and the periodic
boundary conditions that allow for volume change of the domain are
indicated by solid red lines, respectively. b) The implied rectangular
precipitate superstructure. c) and d) Precipitate arrangements \hexx{}
and \hexy{} that arise from point symmetry centers indicated by a
red dot.\label{fig:superstructure}}
\end{figure*}

\fig\ref{fig:superstructure}a) shows a simulation domain with an
aspect ratio of $2.5$ and the imposed boundary conditions for the
displacements normal to the boundaries. Modeling a precipitate in
a periodic configuration implies an arrangement of the precipitates
in a strict long-range order. Specifically, the above boundary conditions
imply a superstructure in which the precipitates are arranged in the
corners of a cuboidal unit-cell \citep{Mueller2000} with the distance
to each nearest neighbor being twice the domain side lengths. Henceforth
this configuration will be referred to as a rectangular superstructure.

By application of antisymmetric variation boundary conditions at one
periodic boundary, one can change the arrangement superstructure \citep{GurevichKarmaPlappTrivedi2010}.
Therefore the tangential strains and the phase-field must be imposed with
a point-symmetric operation with respect to a bisecting point of the
respective domain boundary. \fig\ref{fig:superstructure}b) shows
the rectangular superstructure and \fig\ref{fig:superstructure}c)
and d) show two hexagonal arrangements henceforth referred to as \hexx{}
and \hexy{}. The respective bisecting point that is used for the
antisymmetry operation is indicated as a red dot. Antisymmetric boundary
conditions applied to more than one boundary lead to virtual simulation
domains that include more than one precipitate. This state is not
represented by the actual simulation domain and renders non-physical
solutions.

\section{Simulation results and discussion}

To determine the shape of coherent \gdp{} precipitates we set up
phase-field simulations considering the interfacial energy and elastic
contributions with constant \gdp{} phase fraction. Beginning from
an arbitrary precipitate shape the phase-field converges towards the minimum
of the total energy. The lower left edge of the domain lies in the
precipitates center point with the domain boundaries being parallel
to the crystallographic $\left[100\right]$ and $\left[001\right]$
directions. The observed phase fraction in two dimensions appears
higher than the respective three-dimensional phase fraction. Beginning from
the discussed initial simulation configuration the system was relaxed
for a minimum of $4\cdot10^{5}$ iterations. Subsequently the aspect
ratio and the total energy contributions were evaluated.

To describe the ratio between interfacial and elastic contribution
to the pattern formation we introduce a dimensionless parameter $L$
similar to the one presented in \citep{Voorhees1992} as 
\begin{align}
L= & \frac{lC_{44}\varepsilon_{3}^{2}}{\Gamma},\label{eq:L}
\end{align}
 where $C_{44}\varepsilon_{3}^{2}$ is the elastic energy density
scale with the shear modulus $C_{44}$ of the homogeneous isotropic
elastic data provided in \tab\ref{tab:Elastic-constants} and the
largest misfit strain $\varepsilon_{3}$ from \tab\ref{tab:misfits}
to make it independent of the misfit ratio \ratio{}. $\Gamma$ is
the isotropic interfacial energy density. The length scale of a particle
is defined as $l=\sqrt{Rr}$ and will be used to normalize lengths.
$R$ denotes the major half axes and $r$ the minor half axis of the
precipitate, as shown in \fig\vref{fig:precipitates_scheme}.

\subsection{Variation of anisotropic misfit and elastic constants\label{subsec:Influence-of-anisotropy}}

To evaluate the influence of the anisotropy of the misfit strains
given in \tab\ref{tab:misfits}, the simulation domain was chosen
to be $50\times50$ gridpoints with one quarter of an initial, spherical
particle with a $10$ gridpoint radius in the bottom left corner.
The low phase content together with homogeneous and isotropic elastic
material data leaves only the misfit ratio \ratio{} and the interfacial
energy density $\Gamma$ to determine the equilibrium shape. In this
configuration the tetragonal symmetry of the misfit strain is the
cause of the plate shape of the \gdp{} precipitates. There is significant
uncertainty about the absolute values of the lattice misfit strain
$\varepsilon_{1}$ (see \tab\ref{tab:misfits}). The misfit strain
$\varepsilon_{3}$ is fixed to a value of $30\cdot10^{-3}$ and \ratio{}
is varied to examine the influence of the anisotropy of the misfit
\ratio{} on the precipitate shape. 

\begin{figure*}[tbh]
\begin{centering}
\includegraphics[height=0.66\columnwidth]{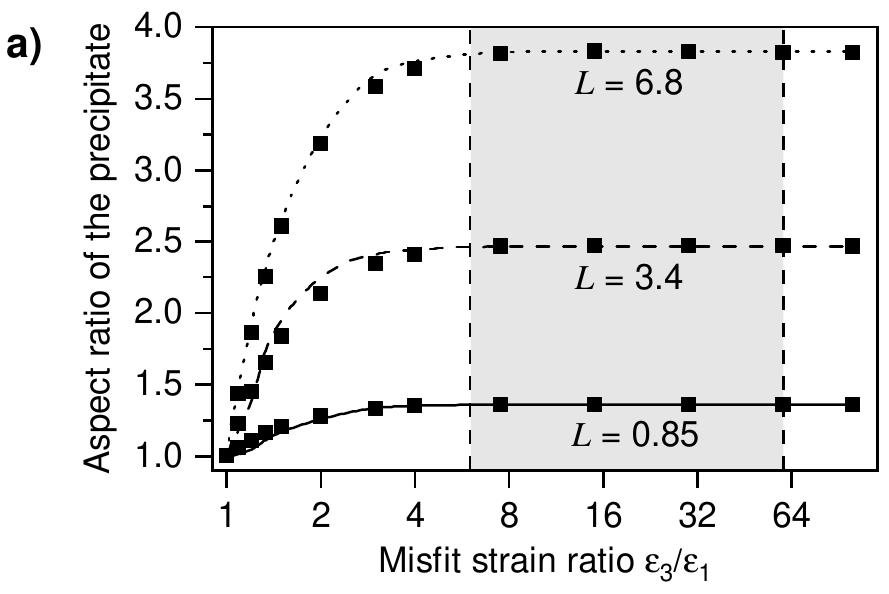}$\quad$\includegraphics[height=0.66\columnwidth]{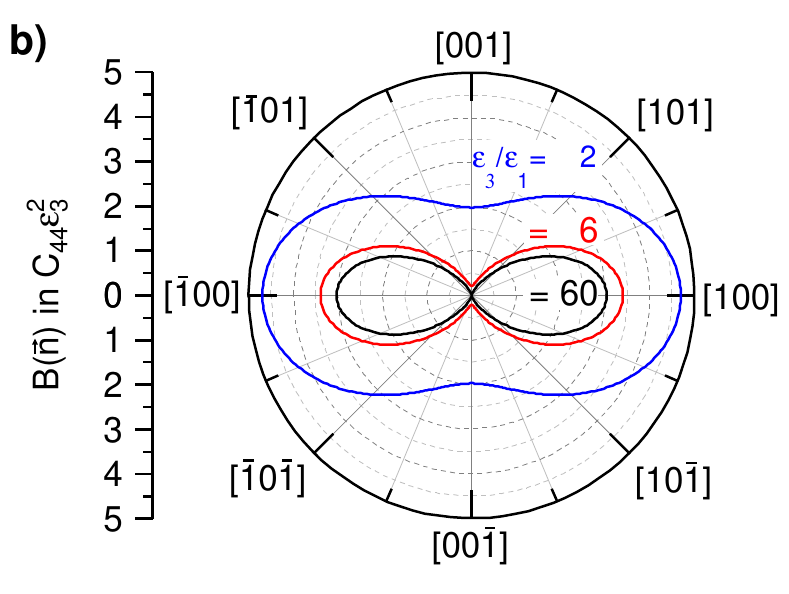}
\par\end{centering}
\caption{a) Precipitate aspect ratio as a function of the misfit ratio \ratio{}
for three different $L$ considering homogeneous and isotropic elasticity
without elastic interaction with neighboring precipitates. The range
in which realistic values for \gdp{} lie is indicated in gray (see
\tab\ref{tab:misfits}). Symbols are phase-field results and the lines
are taken from the model given in \equ(\ref{eq:eshelby1}) and (\ref{eq:eshelby2}).
b) Polar plots of the elastic energy function $B\left(\vec{n}\right)$
for different orientations of the interface normal $\vec{n}$ in the
$\left(010\right)$ plane and different misfit ratios \ratio{}. \label{fig:ratio_study}}
\end{figure*}

\fig\ref{fig:ratio_study}a) shows the precipitate aspect ratio as
determined by the phase-field model as a function of the misfit ratio \ratio{}.
At \ratio{}\,$=1$ the precipitates are circular with an aspect
ratio of $1$. With rising misfit ratio the precipitates show elliptical
shapes with an aspect ratio that rises until a plateau is reached.
In the plateau region, changes in the misfit ratio do no longer influence
the precipitate shape. For all considered values of $L$ the plateau
is reached at \ratio{}$\geq3$. The misfit data considered in \tab\ref{tab:misfits}
lies between \ratio{}\,$=6$ and $60$ and thus lies inside the
plateau region. 

In \fig\ref{fig:ratio_study}a) the simulations are compared to an
analytical model for the optimum shape of \gdp{} precipitates. The
shape is assumed to be a rotational ellipse with an aspect ratio $A$.
The elastic energy density $e$ is homogeneous inside the precipitate
and can be calculated using Eshelby's solution to the inclusion problem
\citep{Eshelby1957} as
\begin{align}
e= & \frac{C_{44}}{1-2\nu}\left(\beta_{1}\varepsilon_{1}^{2}+\beta_{2}\varepsilon_{1}+\beta_{3}\varepsilon_{1}\varepsilon_{3}\right),\label{eq:eshelby1}
\end{align}
where $2\nu=C_{12}/(C_{12}-C_{44})$ and the parameters $\beta_{i}$
only depend on the aspect ratio of the ellipsoid \citep{Cozar1973a,Devaux2008117}.
The optimum aspect ratio is then the aspect ratio that minimizes the
sum of elastic energy and interface energy as follows
\begin{align}
\frac{\partial}{\partial A} & \left(\Gamma S_{0}+eV_{\text{0}}\right)=0,\label{eq:eshelby2}
\end{align}
with $S_{0}(R,A)$ and $V_{0}(R,A)$ being the surface and volume
of an oblate rotational ellipse with major radius $R$ and aspect
ratio $A$ \citep{Cozar1973a}. The aspect ratios determined by the
phase-field model and by the analytical model are very close and the analytic
model shows the same plateau for large misfit ratios.

\fig\ref{fig:ratio_study}b) shows the orientation dependent elastic
relaxation function $B\left(\vec{n}\right)$ that quantifies soft
and hard crystallographic directions independent of the precipitate
shape \citep{khachaturyan1967,morris2010,degeiter2020}. It is defined
as 
\begin{align}
\begin{array}{cccc}
B\left(\vec{n}\right) & = & C_{ijkl}\varepsilon_{ij}^{0}\varepsilon_{kl}^{0} & -\;n_{i}\sigma_{ij}^{0}G_{jk}\left(\vec{n}\right)\sigma_{kl}^{0}n_{l}\\
\sigma_{ij}^{0} & = & C_{ijkl}\varepsilon_{kl}^{0}\\
G_{jk}^{-1}\left(\vec{n}\right) & = & C_{ijkl}n_{i}n_{l}
\end{array}\label{eq:B_n1}
\end{align}
with implied summation over repeated indices. The elastic properties
are assumed to be homogeneous and isotropic (see \tab\ref{tab:Elastic-constants})
and only \ratio{} is varied. For \ratio{}\,$>1$ the direction
minimizing $B\left(\vec{n}\right)$ has been found to be the tetragonal
axis ($\left[001\right]$) \citep{wen1981}. Assuming vanishing interfacial
energy density the equilibrium precipitate shape is an infinitely
extended plate with a ($\left[001\right]$) habit plane \citep{wen1981,morris2010}.
Finite interfacial energy contributes significantly to the formation
of the experimentally observed shapes by counteracting this strong
anisotropy driving force of $B_{\left[100\right]}/B_{\left[001\right]}\gg1$
that is observed for \ratio{}\,$=6$ and $60$. For \ratio{}\,$\gg1$
the precipitate aspect ratio is influenced only by $L$ and the contribution
of $\varepsilon_{1}$ is therefore negligible. In \fig\ref{fig:ratio_study}b),
for \ratio{}\,$=2$ the anisotropy of $B$ is significantly less
pronounced and less interfacial energy density is required to counteract
to that anisotropy to reach a comparable aspect ratio as shown in
\fig\ref{fig:ratio_study}a). 

The two analytical approaches to the tetragonal inclusion problem
discussed above both show, qualitatively and quantitatively, that
for \ratio{}$\gg\infty$ the formation of plate-shaped precipitates
is only driven by $\varepsilon_{3}$. As indicated by the gray area
in \fig\ref{fig:ratio_study}a), the large uncertainty range of the
misfit strain ratio for the \gdp{} phase fully lies within this limit! 

\begin{figure}[tbh]
\begin{centering}
\includegraphics[width=\columnwidth]{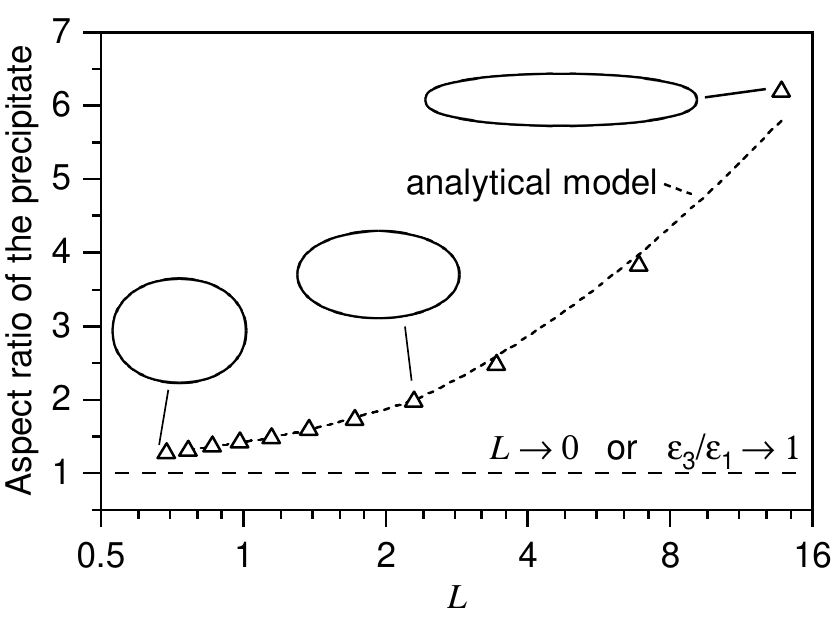}
\par\end{centering}
\caption{Precipitate aspect ratio as a function of $L$ at \ratio{}$=60$
considering isotropic elasticity and no elastic interaction with neighboring
precipitates. Exemplary elliptical precipitate shapes are shown together
with the prediction of the analytical model given in \equ(\ref{eq:eshelby1})
and (\ref{eq:eshelby2}). \label{fig:gamma_study}}
\end{figure}

\fig\ref{fig:gamma_study} shows the aspect ratio of the precipitates
as a function of $L$ for a realistic misfit ratio \ratio{}$=60$.
The aspect ratios of the plate-shaped precipitates increase with increasing
$L$ as the elastic bulk energy of the system dominates the interfacial
part. The precipitate shapes were found to be ellipses with aspect
ratios close to the prediction of \equ(\ref{eq:eshelby1}) and (\ref{eq:eshelby2}).
The good agreement between the two-dimensional phase-field model and the three-dimensional
analytical model in \fig\ref{fig:ratio_study}a) and \ref{fig:gamma_study}
suggests that a two-dimensional model is sufficient to describe the aspect
ratio of a \gdp{} precipitate.

\begin{figure*}[tbh]
\begin{centering}
\includegraphics[height=0.66\columnwidth]{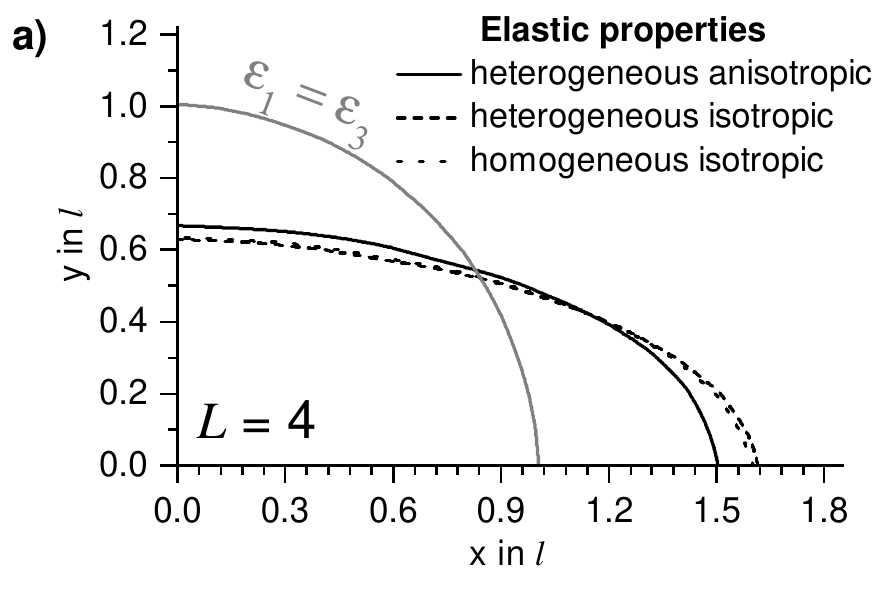}$\quad$\includegraphics[height=0.66\columnwidth]{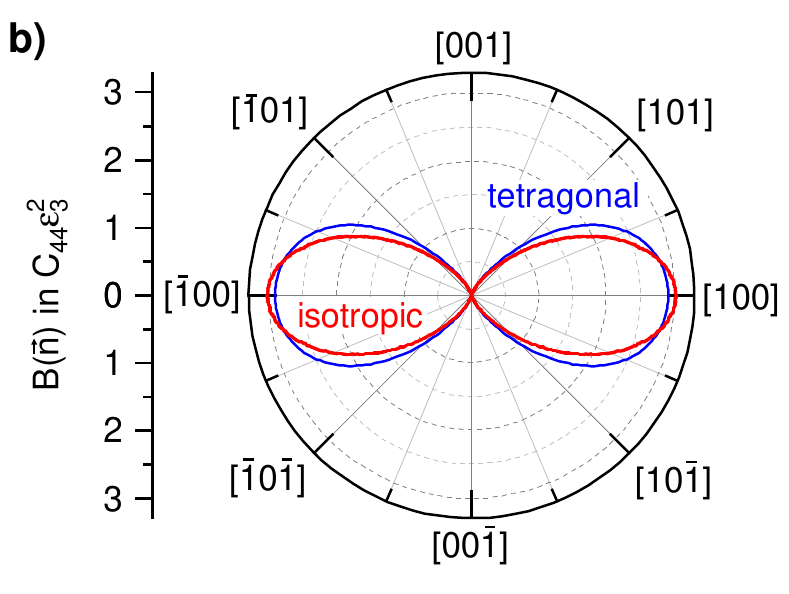}
\par\end{centering}
\caption{a) Influence of inhomogeneity and anisotropy of the elastic properties
on precipitate shapes at $L=4$. Elastic constants are given in \tab\ref{tab:Elastic-constants}.
For reference, the circular shape of a precipitate with $\varepsilon_{1}=\varepsilon_{3}=30\cdot10^{-3}$
and isotropic homogeneous elasticity is shown. b) Polar plots of the
orientation dependent elastic energy function $B\left(\vec{n}\right)$
in the $\left(010\right)$ plane for isotropic and tetragonal elastic
constants\ratio{}.\label{fig:inhomogeneous_anisotropic}}
\end{figure*}

\fig\ref{fig:inhomogeneous_anisotropic}a) shows the influence of
inhomogeneous elastic properties for precipitate and matrix phase
on the precipitate shape as well as the influence of anisotropy of
the elastic constants (see \tab\ref{tab:Elastic-constants}). The
simulation configuration is the same as in \fig\ref{fig:ratio_study}a)
with \ratio{}\,$=60$ and $L=4$. For reference, the circular shape
of a precipitate with an isotropic misfit ($\varepsilon_{1}=\varepsilon_{3}=30\cdot10^{-3}$)
and isotropic homogeneous elasticity is given. In the considered case
inhomogeneous elastic properties do not influence the precipitate
shapes significantly. Anisotropic elastic properties lead to an elliptical
precipitate with a reduced aspect ratio. 

\fig\ref{fig:inhomogeneous_anisotropic}b) shows the elastic energy
density $B\left(\vec{n}\right)$ (see \equ(\ref{eq:B_n1})) for isotropic
and tetragonally anisotropic elastic constants (see \tab\ref{tab:Elastic-constants})
and \ratio{}\,$=60$. The tetragonal anisotropy of the \gdp{} phase
leads to the $\left[100\right]$ direction being elastically softer
than in the isotropic case. For tilted directions orientations $\left\langle 10h\right\rangle $
$B$ changes non-uniformly. The ratio of $B_{\left[100\right]}/B_{\left[001\right]}$
does not change significantly compared to the ratio $B_{\left[100\right]}/B_{\left\langle 10h\right\rangle }$
with $h\neq\left\{ 0,1\right\} $. This qualitative change in the
energetics of tilted interfaces leads to the observed shortening of
the precipitate when anisotropy of the elastic properties is considered.

We conclude that without elastic interaction the formation of plate-shaped
\gdp{} precipitate shapes is mainly driven by the tetragonally anisotropic
misfit in the system. The aspect ratio of the precipitates depends
on the misfit strains and on the interfacial energy density. For realistically
high misfit ratios \ratio{}\,$\gg1$ no influence of the absolute
value of $\varepsilon_{1}$ on the precipitate shape is found. Inhomogeneity
of the elastic constants has negligible influence on the precipitate
shapes. Anisotropic elastic properties lead to elliptical precipitates
with decreased aspect ratios.

\subsection{Precipitate superstructure and particle-particle interaction}

A square simulation domain at realistically high phase contents eventually
leads to precipitates coagulating. Rectangular simulation domains
have to be set up to avoid that. The aspect ratio of the rectangular
simulation domains will be discussed as part of the simulation configuration.
In the following simulations elastic data of the phases is anisotropic,
$L=4$ and \ratio{}\,$=60$ (see \tab\ref{tab:Elastic-constants}
and \ref{tab:misfits}). We describe the interaction between neighboring
precipitates by the particle distance. It is defined as the distance
between the centers of the precipitates in $\left[001\right]$ or
$y$- direction or two times the height of the simulation domain.

\begin{figure}[tbh]
\begin{centering}
\includegraphics[width=\columnwidth]{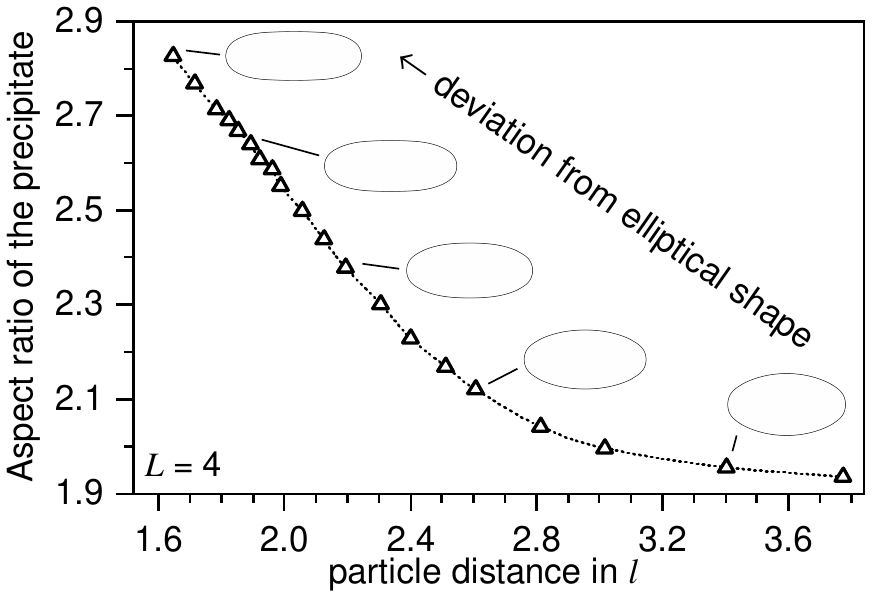}
\par\end{centering}
\caption{Dependence of the precipitate aspect ratio on the particle distance
in $y$-direction with exemplary precipitate shapes. The aspect ratio
of the simulation domain is $2.5$. With increasing \gdp{} volume
content, the aspect ratio increases and the precipitates deviate from
their elliptical shape.\label{fig:phi_study}}
\end{figure}

\fig\ref{fig:phi_study} shows the results of a simulation study
in a domain with a fixed aspect ratio of $2.5$ and rectangular superstructure.
The size of the initial elliptical precipitate was kept constant ($R=100$
and $r=34$ gridpoints) and the relative size of the simulation domain
was varied in order to model different particle distances. The aspect
ratio of a particle increases by $\unit[35]{\%}$ in a system when
the particle distance is reduced from $3.8l$ to $1.6l$. Note that
due to the conserved aspect ratio of the domain also the particle
distance in $x$-direction is reduced simultaneously and the phase
fraction is rising. Interestingly, the increase in the aspect ratio
at high phase contents does not lead to a simple stretching of the
particle but also to a deviation from the elliptical shape. The precipitate
shape has reduced curvature along its major extent. Similar elastic
interactions affecting the precipitate shape have been reported for
\g{}/\gp{} with volume fractions up to $\unit[75]{\%}$ \citep{Mueller2000}.
Distant precipitates experience attraction and close precipitates
are repulsed leading to an equilibrium matrix channel width between
the precipitates \citep{Su1996,Su1996a,GoerlerLopezGalileaRonceryShchygloTheisenSteinbach2017,Jokisaari2017}.

\begin{figure}[tbh]
\begin{centering}
\includegraphics[width=\columnwidth]{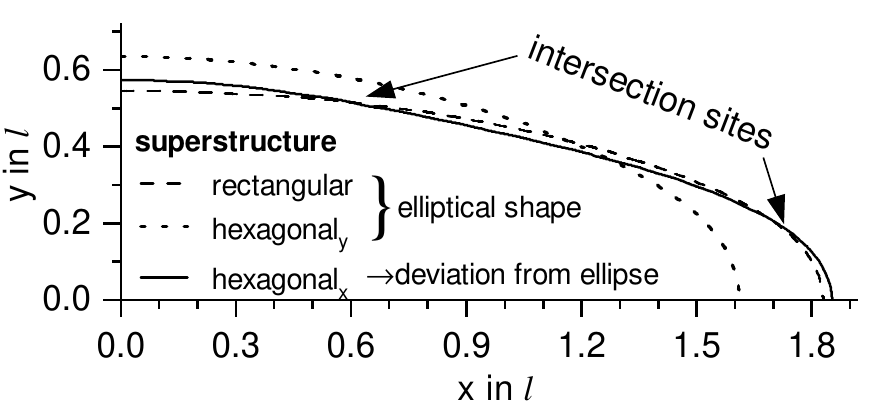}
\par\end{centering}
\caption{Equilibrium shapes of precipitates subject to different boundary conditions
implying three superstructures. The superstructure nomenclature is
introduced in \fig\ref{fig:superstructure}. Arrows indicate intersections
of the elliptical shape outline (rectangular) with the non-elliptical
shape (\hexx).\label{fig:superstructure-shapes}}
\end{figure}

\fig\ref{fig:superstructure-shapes} shows precipitate shapes influenced
by elastic particle-particle interaction in different superstructures
at a particle distance of $1.8l$. The determined precipitate shapes
subject to different implicit superstructures illustrate the significant
influence of long-range order on precipitate shapes. The area of all
three shapes is equal. The shapes for rectangular and \hexy{} arrangement
can be approximated with ellipses. The superstructure \hexy{} shortens
the precipitate by $\unit[10]{\%}$ but it remains an ellipse. The
\hexx{} arrangement leads to deviation from the elliptical shape
visible due to the fact that the phase boundary of the precipitate
``rectangular'' is intersected twice by the phase boundary of ``\hexx''.
The sites of the intersections are indicated by arrows.

\begin{figure}[tbh]
\begin{centering}
\includegraphics[width=\columnwidth]{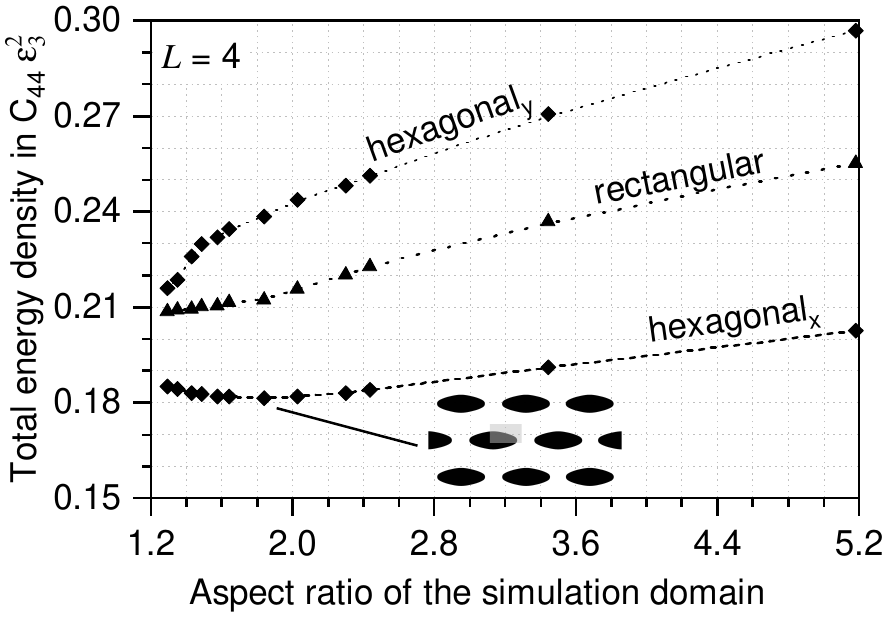}
\par\end{centering}
\caption{Mean energy densities of systems plotted against the domain aspect
ration. Results are shown for a particle distance of $1.8l$ and three
different superstructures each. The superstructure nomenclature is
introduced in \fig\ref{fig:superstructure}. The energetically most
favorable microstructure is plotted.\label{fig:E_study}}
\end{figure}

\fig\ref{fig:E_study} shows the total energy density plotted against
the domain aspect ratio for the three possible superstructures at
a particle distance of $1.8l$. The \hexx{} arrangement is the energetically
most favorable. The rectangular arrangement is intermediate and the
\hexy{} superstructure shows the highest energy density. The rectangular
and the \hexy{} superstructure both exhibit a steadily dropping energy
density for aspect ratios of the simulation domain close to 1. It
is only in the energetically most favorable \hexx{} arrangement that
one finds a distinct minimum at an aspect ratio of the simulation
domain of $1.8$. The implied microstructure of such an energetically
optimum configuration is also given in \fig\ref{fig:E_study}. For
spherical precipitates with an isotropic misfit a cubic arrangement
was found to be the energetic optimum \citep{khachaturyan1974spatially}.

The configuration that was found to be the energetically most favorable
is the one where precipitates exhibit the largest distance to their
nearest neighbors in the direction of the highest misfit strain $\varepsilon_{3}$.
An effect of the high elastic energy contribution in the \hexy{}
arrangement is visible in \fig\ref{fig:superstructure-shapes}, where
it leads to additional shortening of the precipitate. Being the energetically
minimum configuration of the presented model a \hexx{} superstructure
with a domain aspect ratio of $1.8$ will be used to describe a realistic
single-variant \g{}/\gdp{} microstructure with a realistic volume
fraction. 

\subsection{Energy density of the \g{}/\gdp{} interface}

To reproduce experimentally observed aspect ratios of \gdp{} precipitates,
we set up a simulation study assuming constant misfit strains and
isotropic interfacial energy density. The simulation domain has an
aspect ratio of $1.8$, the initial particle has a radius of $20$
gridpoints and a \hexx{} superstructure was used. The sizes of the
respective simulation domains were set such that they resemble a realistic
\gdp{} volume fraction of $\unit[12]{\%}$. We approximate the volume
fraction as the ratio between the volume of a spheroid with the same
aspect ratio as the two-dimensional shape and the volume of a respective
three-dimensional simulation domain by exploiting the system's four-fold
rotational symmetry around the tetragonal axis. The approximated volume
content $\widetilde{V}$ is then given by
\begin{align}
\widetilde{V}= & \frac{4\pi}{3Ad^{3}\sqrt{1.8}},\label{eq:volume_content}
\end{align}
where $A$ is the aspect ratio of the precipitate and $d$ is the
particle distance normalized by $l$. As shown in \fig\ref{fig:superstructure-shapes}
the precipitate shapes deviate from the elliptical shape and therefore
assuming a spheroidal precipitate as in \equ(\ref{eq:volume_content})
is not accurate. We assume that this inaccuracy is negligible as the
inaccuracy of the volume content measurement itself is comparably
large. Note that the single-variant microstructure discussed in this
work reflects material aged under load for several hours \citep{Gao1996,Zhou2014270}
or specifically tailored material \citep{Zhang2019}. We assume that
particle interactions of perpendicularly oriented precipitates lead
to the same stretching of the precipitates but might impose further
constraint to the growth of precipitates that is not reflected by
the proposed model. 

\begin{figure*}[tbh]
\begin{centering}
\includegraphics[width=0.95\columnwidth]{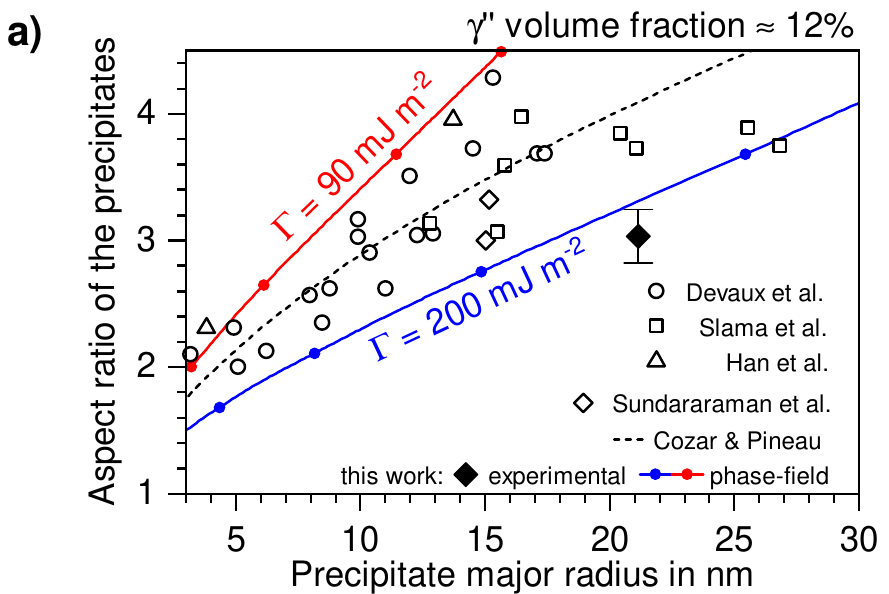}\includegraphics[width=0.95\columnwidth]{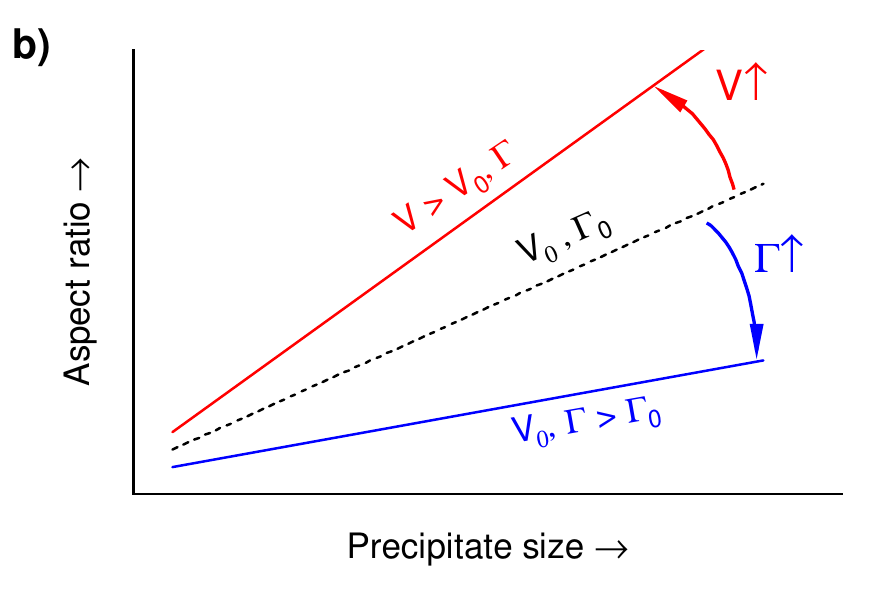}
\par\end{centering}
\caption{a) Data set of precipitate aspect ratio over major radius in the regime
of fully coherent interfaces \citep{Han1982,sundararaman.92,slama1997,Devaux2008117}.
Simulation-based estimates for a lower and upper boundary of the interfacial
energy density are given for a \gdp{} volume fraction of $\unit[12]{\%}$.
The prediction of the size dependent aspect ratio of \gdp{} precipitates
with an interfacial energy of $\unit[90]{\mJm}$ by the analytical
model of Cozar \& Pineau is also shown \citep{Cozar1973a}. b) Schematic
plot illustrating the competing effects of volume fraction $V$ and
interfacial energy density $\Gamma$ on the precipitate aspect ratio.\label{fig:data_study}}
\end{figure*}

\fig\ref{fig:data_study}a) shows experimental data from various
sources reporting the aspect ratio of \gdp{} precipitates as a function
of the precipitate major radius \citep{Han1982,sundararaman.92,slama1997,Devaux2008117}.
Interfacial dislocations occur when full coherency of precipitates
larger than $\unit[R=25]{nm}$ is lost \citep{Cozar1973a,slama1997,Devaux2008117}.
Here, we restrict to fully coherent precipitates, as interfacial dislocations
alter the strain field around a precipitate \citep{JiLouRowattZhangSimpsonChen2016}
in a way that is not reflected by the model. The aspect ratio of the
precipitates increase with increasing precipitate size due to the
rising importance of elastic bulk effects over the interfacial energy
with $L\sim l$ (see \equ(\ref{eq:L})). In simulations with a particle
distance of $5l$ an interfacial energy density of $\unit[65]{\mJm}$
to $\unit[130]{\mJm}$ is necessary to reproduce experimentally observed
aspect ratios. This is in good accordance with the previously reported
isotropic \g{}/\gdp{} interfacial energy density of $\unit[95\pm17]{\mJm}$
\citep{Devaux2008117}. This value was obtained by the application
of an analytical model for the equilibrium aspect ratio of a \gdp{}
precipitate given in \equ(\ref{eq:eshelby1}) and (\ref{eq:eshelby2})
based on Eshelby's inclusion theory \citep{Eshelby1957,Cozar1973a}.
This model assumes isotropic and homogeneous elasticity, isotropic
interfacial energy density, a spheroidal precipitate shape and no
elastic interaction. The prediction of this model for the size dependent
aspect ratio of a precipitate with an interfacial energy density of
$\unit[90]{\mJm}$ is also shown in \fig{}\ref{fig:data_study}a).
An interfacial energy density of $\unit[145]{\mJm}$ was found for
$\mathrm{D}0_{22}$ precipitates in a Fe-Ni-Ta alloy \citep{Cozar1973a}.

\fig\ref{fig:data_study}a) shows that at a realistic volume fraction
of $\unit[12]{\%}$ the isotropic energy density needed to reproduce
experimental findings lies between $\unit[90]{\mJm}$ and $\unit[200]{\mJm}$,
which corresponds to $0.8<L<5.4$. In a real system the interfacial
energy of a tetragonal to cubic interface might be strongly anisotropic
\citep{vaithyanathan2004,kim2017}, which could also be included in
a phase-field model. As to the best of our knowledge, no information about
the magnitude of such an anisotropy for the coherent \g{}/\gdp{}-interface
is available, we assume an isotropic $\Gamma$. The obtained isotropic
interfacial energy densities are a factor of $1.5$ higher than those
determined without consideration of elastic particle-particle interaction.
\fig\ref{fig:data_study}a) includes data from averaged aspect ratios
generated from TEM images as for example given in \fig\ref{fig:SEM}b).
72 fully coherent \gdp{} precipitates in IN718M were evaluated
by image analysis. The samples were homogenized at $\unit[1423]{K}$
for $\unit[2]{h}$ and subsequently water quenched. The error bar
shows the standard deviation of the aspect ratio. For this system,
we predict an isotropic interfacial energy density of $\unit[220]{\mJm}$. 

\fig\ref{fig:data_study}b) illustrates the influence of the competing
factors discussed above on the aspect ratio of a precipitate. A higher
the interfacial energy density $\Gamma$ provides a tendency towards
more spherical precipitates, i.e.~smaller aspect ratios. Stronger
elastic interactions between precipitates at higher volume fraction
$V$, in turn, lead to higher precipitate aspect ratios (see \fig{}\ref{fig:data_study}a).
To estimate the interfacial energy density from the experimentally
observed aspect ratios of precipitates it is crucial to take into
account the elastic interactions of precipitates at finite volume
fractions.

Comparison of theoretically determined equilibrium shapes with experimentally
observed shapes is a possible way to get information about the interfacial
energy \citep{lanteri1986morphology,Devaux2008117,Holzinger2019}.
However, the accuracy of this method is limited by the underlying
model description of the equilibrium shape and by the possibility
to experimentally observe precipitates in their equilibrium. The model
presented in this work is in many ways an improvement over existing
analytical models as it takes into account elastic interactions between
precipitates in an two-dimensional  optimum, non-rectangular arrangement and
tetragonal/cubic anisotropy of the elastic constants for both phases.
It is limited by being only a realistic description of a uniform single-variant
microstructure that does not take into account the kinetics of precipitate
growth that might have strong influence on the shapes of experimentally
observed precipitates \citep{vaithyanathan2002PRL}. It was also found
that a periodic arrangement of precipitates must not always be a stable
configuration \citep{degeiter2020}.

\section{Conclusion}

We evaluate influencing factors on equilibrium shapes of \gdp{} precipitates
in Ni-based superalloys considering one orientational variant. The shapes
are determined using a phase-field formulation taking into account interfacial
and elastic energy contributions.
\begin{enumerate}
\item At negligible elastic particle-particle interaction, the phase-field model
provides elliptic equilibrium shapes \gdp{} that are fully consistent
with former analytic descriptions \citep{Eshelby1957,Cozar1973a}.
 The aspect ratio $R/r$ of a precipitate increases with increasing
misfit strain $\varepsilon_{3}$. The precipitate shape is influenced
by $\varepsilon_{1}$ only for \ratio{}\,$\leq3$. Inhomogeneity
and anisotropy of the elastic constants have less significant influence
on the \gdp{} precipitate shape.
\item Elastic particle-particle interactions significantly influence precipitate
shapes at realistically high volume fractions. Different periodic
arrangements of \gdp{} precipitates are modeled by respectively tailored
boundary conditions. A decreased particle distance leads to an increased
precipitate aspect ratio.
\item Non-volume conserving displacement boundary conditions allow precise
determination of the total energy density of periodic precipitation
microstructures. The energetically most favorable superstructure is
a \hexx{} precipitate arrangement (see \fig\ref{fig:superstructure}c). 
\item The evaluation of interfacial energy density based on equilibrium
shapes is sensitive to the phase content. At a realistic volume fraction
of $\unit[12]{\%}$, an interfacial energy density between $90$ and
$\unit[200]{\mJm}$ leads to precipitate aspect ratios that match
experimental observations. Respective interfacial energy densities
determined without accounting for elastic interaction between the
precipitates are $\unit[30]{\%}$ lower.
\end{enumerate}

\section*{CRediT authorship contribution statement}

\textbf{F. Schleifer:} Conceptualization, Methodology, Software,
Validation, Formal Analysis, Investigation, Writing -- Original Draft,
Visualization. \textbf{M. Holzinger:} Conceptualization, Methodology,
Software. \textbf{Y.-Y. Lin:} Conceptualization, Investigation, Data
Curation, Validation. \textbf{U. Glatzel:} Supervision, Resources,
Writing -- Review \& Editing, Project Administration, Funding Acquisition.
\textbf{M. Fleck:} Conceptualization, Software, Supervision, Writing
-- Review \& Editing, Project Administration, Funding Acquisition.

\section*{Declaration of Competing Interest}

Authors have no conflict of interest to declare.

\section*{Acknowledgements}

This work is funded by the Deutsche Forschungsgemeinschaft (DFG) in
the priority program SPP 1713 (GL181/53-1|FL826/3-1). We thank the
Federal Ministry of Education and Research (BMBF) for the financial
support under the running project ParaPhase (funding code: 01IH15005B).
The financial support of the Federal Ministry for Economics and Energy
(BMWi) of the Federal Republic of Germany under the running project
COORETEC: ISar (funding code: 03ET7047D) is greatly acknowledged.
We thank our former colleague F. Krieg for the provision of RUS data
and A. Finel from ONERA in Châtillon, France for the discussion on
tetragonal anisotropy.

\bibliographystyle{model1-num-names}
\addcontentsline{toc}{section}{\refname}\bibliography{LiteratureGammaDP}

\end{document}